\documentclass[a4paper,11pt]{article}
\usepackage{lineno}
\usepackage{pos}
\graphicspath{
  {IMAGES/}
}

\title{Clocking the particle production and tracking of strangeness balance and radial flow effects at top LHC energy with ALICE}
\ShortTitle{Clocking the particle production and tracking quantum number balancing}

\author*[a]{Victor Gonzalez}
\onbehalf{for the ALICE collaboration}

\affiliation[a]{Department of Physics and Astronomy, Wayne State University,\\
  666 W Hancock, Detroit, USA}

\emailAdd{victor.gonzalez.sebastian@gmail.com}

\abstract{Balance functions have been used extensively to elucidate the time evolution of quark production in heavy-ion collisions. Early models predicted two stages of quark production, one for light quarks and one for the heavier strange quark, separated by a period of isentropic expansion. This led to the notion of clocking particle production and tracking radial flow effects, which drive the expansion of the system. In this talk, balance functions of identified particles in different multiplicity classes of pp collisions at $\sqrt{s} = 13.6\;\text{TeV}$ recorded by ALICE during the LHC Run 3 are reported. The results are compared with different models as well as with previously published results on pp and Pb--Pb collisions at different energies. The results enable tracking the balancing of electric charge and strangeness by measuring how the widths and integrals of the charge and strangeness balance functions evolve across different collision energies.}

\FullConference{42nd International Conference on High Energy Physics (ICHEP2024)\\
18-24 July 2024\\
Prague, Czech Republic\\}

\begin{document}
\renewcommand{\logo}{\relax}
\maketitle

\section{Introduction}

When addressing high-energy hadronic collisions, quantum numbers are expected to be conserved.
As such, particles are created, correlated, in pairs. With further evolution, the pairs keep
correlated. In an expansion scenario the larger the pair lifetime the longer the correlation reach.
In a perfect full acceptance detector the different quantum numbers will be fully balanced.
Actual detectors put limits to the amount of balancing able to be measured but, no doubt,
such a measure is of the greatest interest to unveil mechanisms of particle production and 
transport.

\section{Charge balance function}

Charge balance function (BF) was initially proposed as tool for probing radial flow and unveiling the timing of, clocking, the particle production~\cite{Bass:2000az}. 
Naively, in a rapidly expanding scenario, particle pairs tend to be closer the stronger is the expansion. Otherwise particles pairs created at different times in the evolution of the system are expected to have different reach in their correlations. 

The suggested expression of the charge BF~\cite{Bass:2000az},
\begin{equation}
  B = \frac{1}{2}\left[
        \frac{N^{+-}}{N^{+}} - \frac{N^{--}}{N^{-}} + \frac{N^{-+}}{N^{-}} - \frac{N^{++}}{N^{+}}
    \right],
    \label{eq:bforig}
\end{equation}
involves the difference between opposite charge per particle and same charge per particle terms. 
The amount of balanced charge is given by the integral of the BF while the reach of the balancing is inferred from their widths. 
The evolution on different expansion scenarios can be addressed by extracting the charge BF from collisions of a wide set of multiplicity or centrality classes. 
Due to the impact of the limited acceptance, results with different acceptances were able to be compared on the same foot only after considering the techniques described in Ref.~\cite{STAR:2010slb}.

Charge BFs from unidentified particles in pp collisions at $\sqrt{s} = 7\;\text{TeV}$, and p--Pb and Pb--Pb collisions at $\sqrt{s_{\rm NN}}$ = 5.02 and 2.76 TeV, respectively, were published by the ALICE collaboration~\cite{ALICE:2015nuz}. 
Longitudinal and azimuthal BF widths were extracted out of BF from particles at three transverse momentum ($p_{\rm T}$) ranges, low, intermediate, and high $p_{\rm T}$.
The overall narrowing towards higher multiplicity collisions along the three systems in the bulk regime,
low $p_{\rm T}$ range, showed a scenario dominated by a radial expansion. The soft transition in the widths evolution
from pp to p--Pb pointed to a similar mechanism for particle production in both systems. The discontinuous
transition to Pb--Pb potentially sets in place a different driver. Otherwise, at intermediate and high $p_{\rm T}$
signs of the same mechanism prevail along the three systems. 

Charge BFs from identified $\pi$ and $\rm K$ in Au--Au and d--Au collisions at $\sqrt{s_{\rm NN}} = 200\;\text{GeV}$, and pp collisions at $\sqrt{s} = 200\;\text{GeV}$ were published by the STAR collaboration~\cite{STAR:2010plm} while
charge BFs from identified $\pi$, $\rm K$, and $\rm p$ in Pb--Pb collisions at $\sqrt{s_{\rm NN}} = 2.76\;\text{TeV}$ were published by the ALICE collaboration~\cite{ALICE:2021hjb}. The results from
both experiments are consistent with a radial flow expansion scenario showing signs of two stages of particle production.

Generalized balance functions are inferred from Eq.~(\ref{eq:bforig}) by considering other quantum numbers instead of the electric charge~\cite{Pruneau:2022mui}. For the longitudinal acceptance chosen for this work, pseudorapidity $|\eta|<0.8$, two-particle differential generalized BFs are defined as  
\begin{equation}
    B^{\alpha \beta}\left( \Delta \eta, \Delta \varphi \right) 
        = \frac{1}{2} \left\{
            \rho_1^{\bar{\beta}} \left[
                R_2^{\alpha \bar{\beta}} 
                  - {R_2^{\bar{\alpha} \bar{\beta}}} 
            \right] +
            \rho_1^{\beta} \left[ 
                R_2^{\bar{\alpha} \beta}
                  - {R_2^{\alpha \beta}} 
            \right] 
    \right\},
    \label{eq:bfgen}
\end{equation}
where $\alpha$ and $\beta$ represent realization of the quantum numbers of interest, $\rm p$ and $\rm \bar{p}$, for instance, for baryon and antibaryon, respectively.
$R_{2}$ is the normalized second order cumulant defined as
\begin{align*}
  R_2^{\alpha\beta} \left( \Delta \eta, \Delta \varphi \right)  
    &= \frac{\rho_{2}^{\alpha\beta}} 
        {\rho_{1}^{\alpha} \rho_{1}^{\beta} } 
        - 1,
\end{align*}
with $\rho_{1}^{\alpha} = {\rm d}^{2} N^{\alpha}/{\rm d}\eta\,{\rm d}\varphi$ and 
$\rho_{2}^{\alpha\beta} = {\rm d}^{2} N^{\alpha\beta}/{\rm d}\Delta\eta\,{\rm d}\Delta\varphi$ the single- and two-particle densities, and with $\Delta\eta$ and $\Delta\varphi$ denoting the separation in pseudorapitity and azimuthal angle $\varphi$.
The use of Eq.~(\ref{eq:bfgen}) automatically compensates the BF for the limited acceptance.

\section{Analysis conditions}

The results presented in this work correspond to 70 millions selected events measured by the ALICE detector during one of the first periods of the pp 2022 campaign of the LHC Run 3. 
The events were selected within 7 cm range of the center of the detector along the beam axis.
Event multiplicity was estimated based on the signal measured by Cherenkov arrays on both sides of the interaction point. Events were classified in ten multiplicity classes from 0--10\% (highest multiplicity) to 90--100\% (lowest multiplicity) of the pp interaction cross-section. 
Tracks were reconstructed using information from the silicon based inner tracking system (ITS) and the time projection chamber (TPC). Tracks were required to have a short distance of closer approach to the interaction vertex for reducing contamination from secondaries and weak decays. Only tracks within the acceptance $|\eta|<0.8$ and with $p_{\rm T}$ in the particle production bulk regime, $0.2 < p_{\rm T} < 2\;\text{GeV}/c$, were selected for analysis.

\section{Results and discussion}
\begin{figure}[ht]
    \includegraphics
    [width=0.32\textwidth,keepaspectratio=true,clip=true,trim=00pt 0pt 50pt 0pt]
    {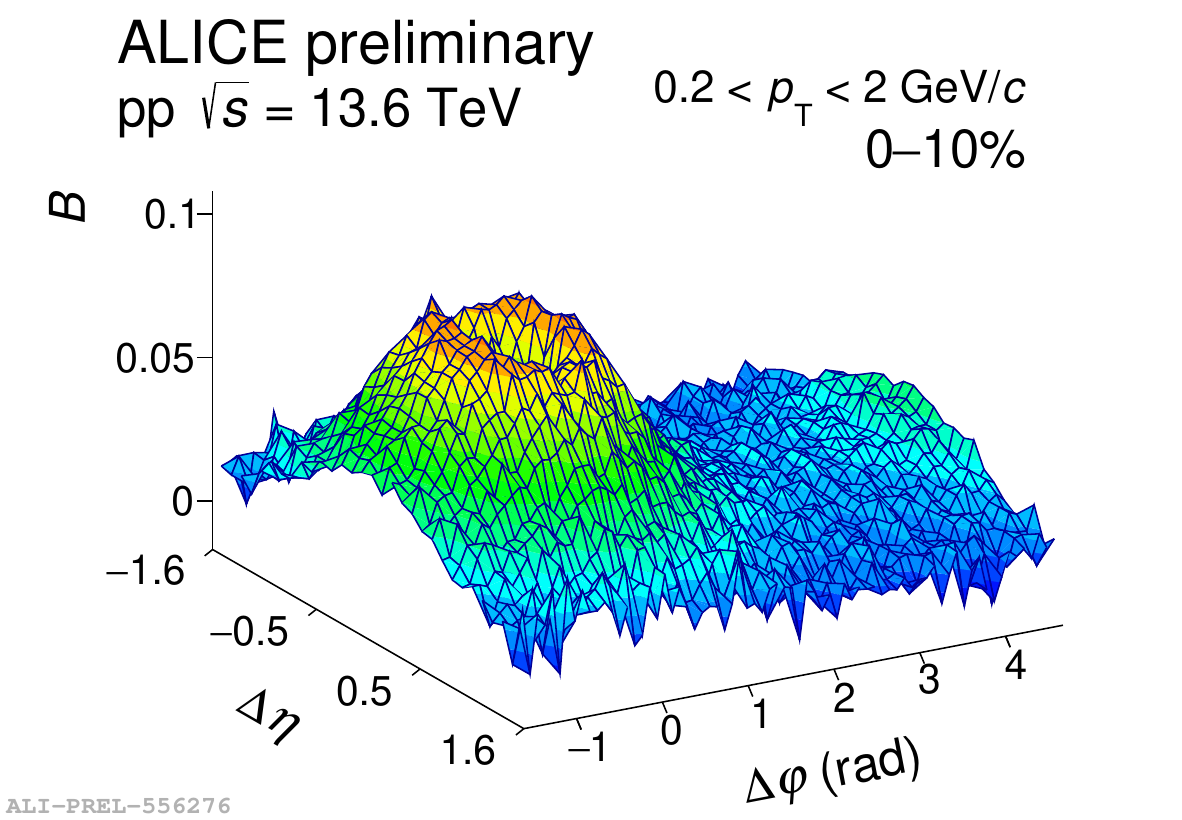}
    \includegraphics
    [width=0.32\textwidth,keepaspectratio=true,clip=true,trim=00pt 0pt 50pt 0pt]
    {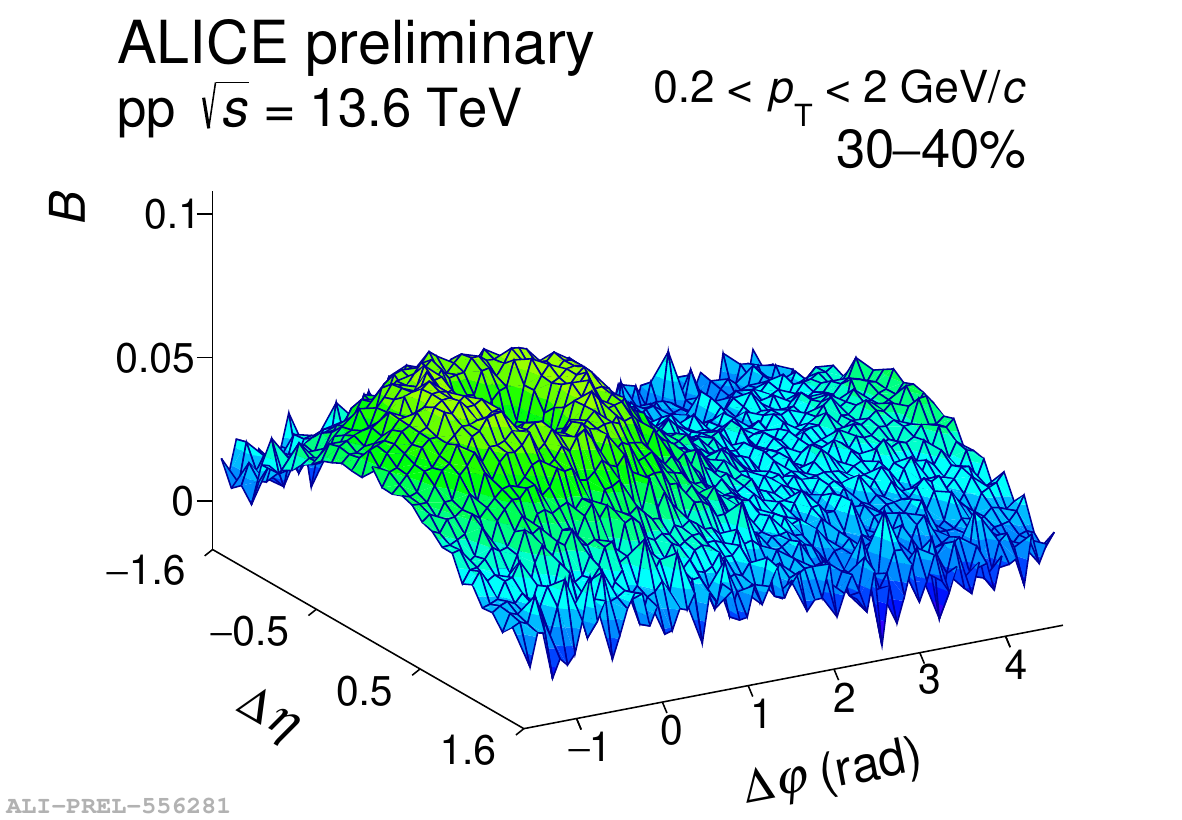}
    \includegraphics
    [width=0.32\textwidth,keepaspectratio=true,clip=true,trim=00pt 0pt 50pt 0pt]
    {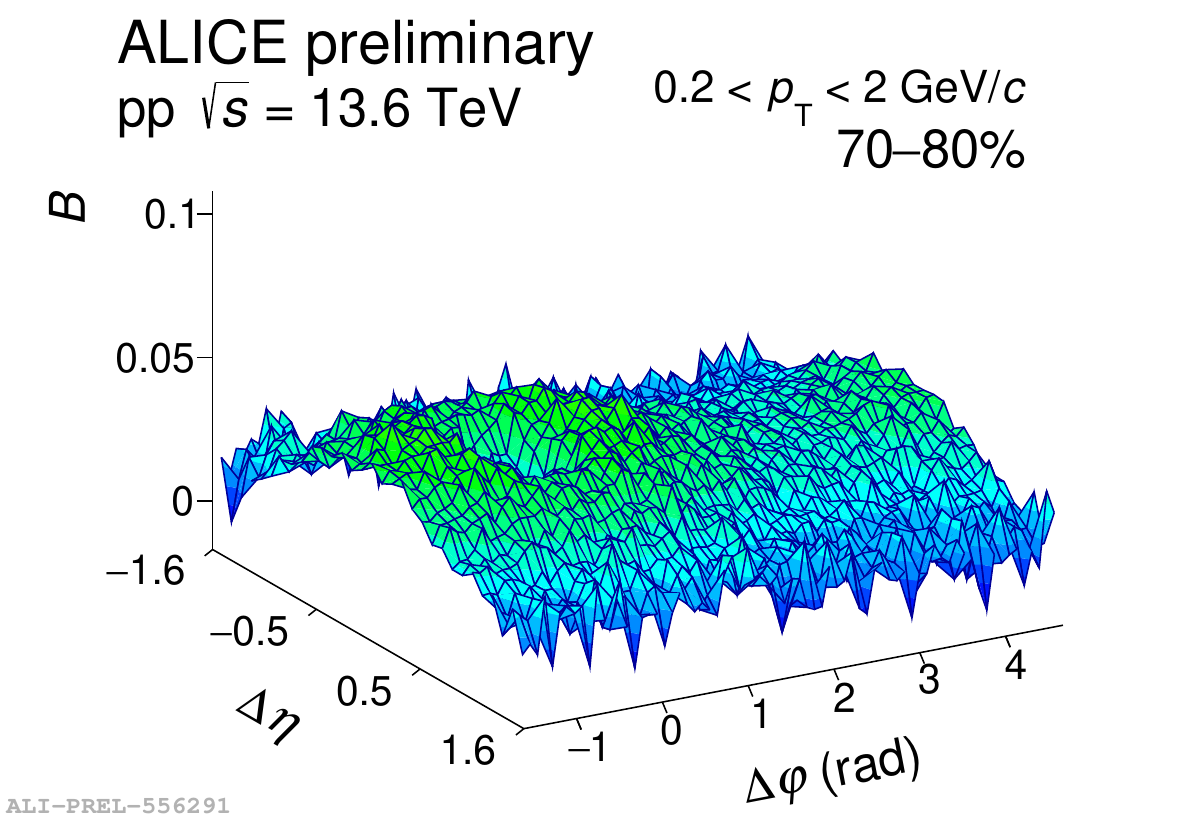}
    \caption{Differential charged hadrons balance functions from 0--10\% (left), 30--40\% (center),
    70--80\% (right), multiplicity classes as measured in pp collisions at $\sqrt{s} = 13.6\;\text{TeV}$.}
    \label{fig:bf2d}
\end{figure}

Differential charged hadron balance functions in selected multiplicity classes of pp collisions at $\sqrt{s} = 13.6\;\text{TeV}$ are shown in Fig.~\ref{fig:bf2d}. 
The prominent near-side peak in high-multiplicity events shows that charge balancing mainly occurs in proximity which can be interpreted as a signature that particles that are correlated stay correlated and focused. With decreasing multiplicity the balancing gets distributed along the azimuthal dimension which can be interpreted as particle evolution not allowing correlated particles to remain focused. The hole on the center of the near-side peak is most probably due to correlations from quantum statistics effects on the same sign component of the BF from pairs with low momentum difference.  

\begin{figure}[hb]
    \includegraphics
    [width=0.48\textwidth,keepaspectratio=true,clip=true,trim=00pt 0pt 50pt 0pt]
    {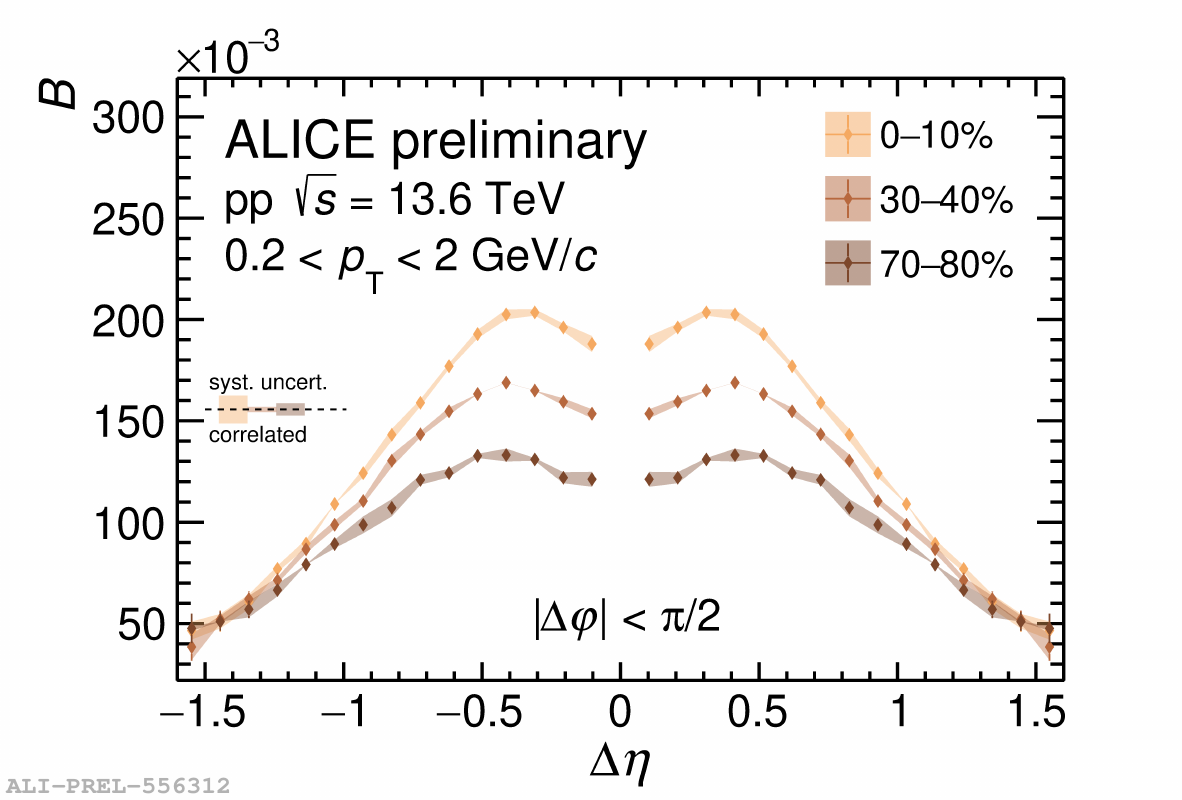}
    \includegraphics
    [width=0.48\textwidth,keepaspectratio=true,clip=true,trim=00pt 0pt 50pt 0pt]
    {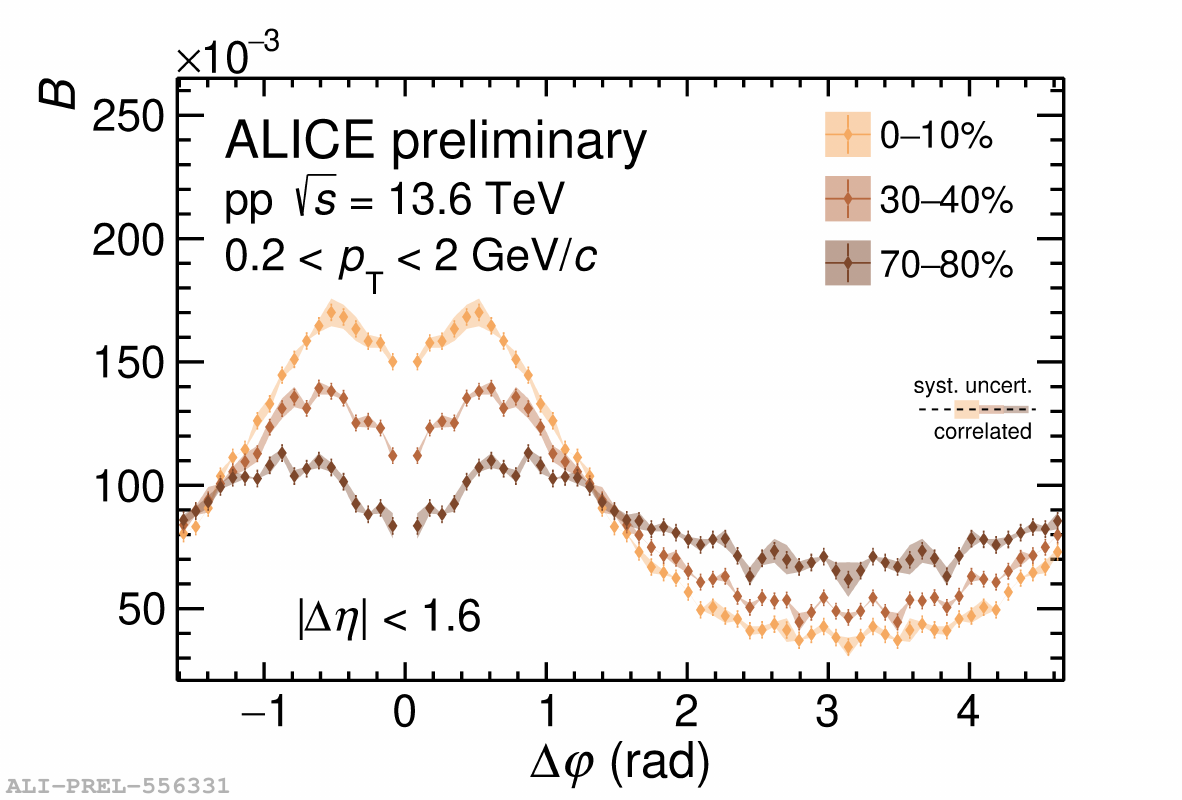}
    \caption{Longitudinal near-side (left) and azimuthal (right) projections of the differential charged hadrons balance functions from selected multiplicity classes as measured in pp collisions at $\sqrt{s} = 13.6\;\text{TeV}$. Vertical bars and shadowed bands represent statistical and uncorrelated systematic uncertainties, respectively, while small boxes along the small horizontal dotted line represent bin by bin correlated systematic uncertainty.}
    \label{fig:bfprojections}
\end{figure}

To further interpret the evolution with multiplicity, Fig.~\ref{fig:bfprojections} shows the longitudinal and azimuthal projections of the charged hadron balance functions for selected multiplicity classes. The redistribution of the charge balancing from the near-side towards the whole azimuthal coverage with decreasing multiplicity is better illustrated here.
Furthermore, it would be tempting to interpret the apparent enlargement of the near-side hole with decreasing multiplicity as a reduction in the size of source of the particle production.
The reach of particle correlation, i.e. charge balancing, which comes from the production in pairs and further evolves, is inferred by studying the widths of the BF. 
\begin{figure}[ht]
    \includegraphics
    [width=0.32\textwidth,keepaspectratio=true,clip=true,trim=00pt 0pt 50pt 30pt]
    {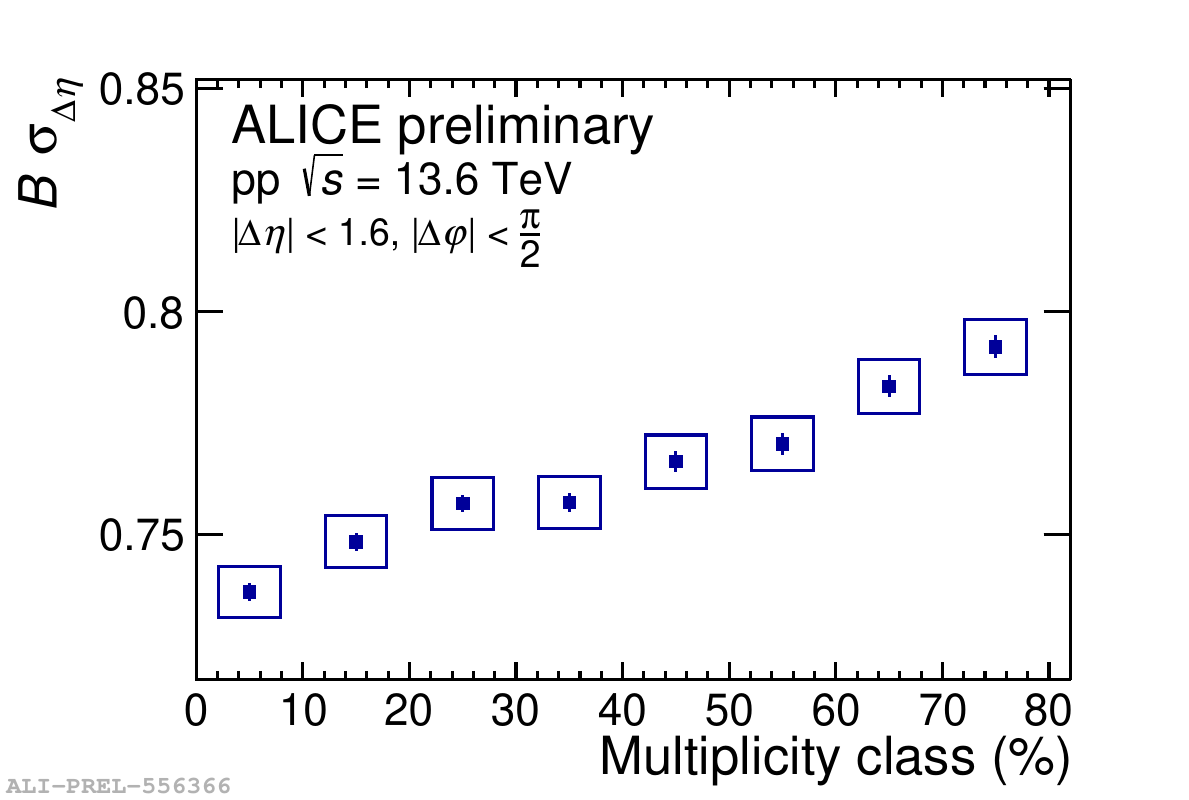}
    \includegraphics
    [width=0.32\textwidth,keepaspectratio=true,clip=true,trim=00pt 0pt 50pt 30pt]
    {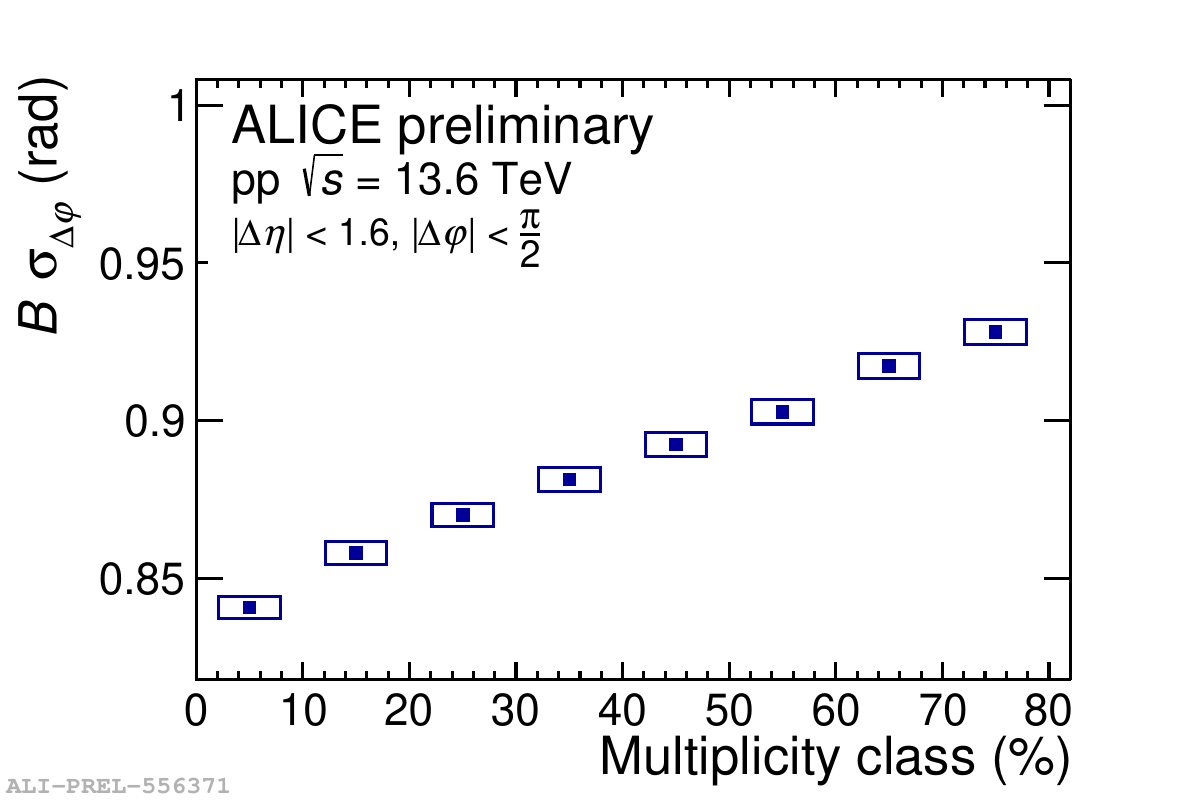}
    \includegraphics
    [width=0.32\textwidth,keepaspectratio=true,clip=true,trim=00pt 0pt 50pt 30pt]
    {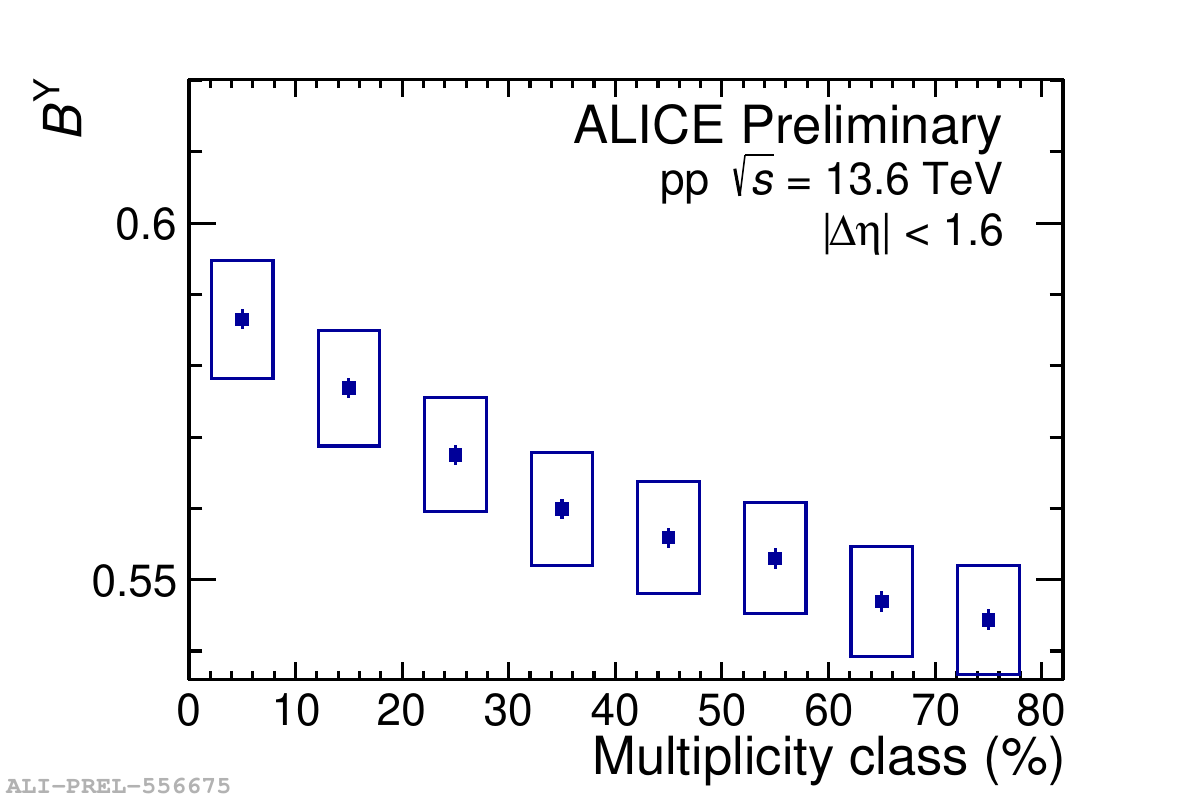}
    \caption{Evolution with multiplicity of the charged hadrons balance functions longitudinal (left) and azimuthal (center) widths, and their integrals (right), from pp collisions at $\sqrt{s} = 13.6\;\text{TeV}$. Vertical bars and rectangular boxes represent statistical and systematic uncertainties, respectively.}
    \label{fig:widhtsandintegrals}
\end{figure}
The evolution with multiplicity of the longitudinal and azimuthal widths of the charged hadrons balance function are shown in the left and center panels of Fig.~\ref{fig:widhtsandintegrals}, respectively. 
The widths evolution with multiplicity is consistent with an expansion scenario and compatible with previous observations~\cite{ALICE:2015nuz}. 
Scenarios with higher momentum, higher multiplicities, keep correlated particles focused and drive narrower BFs.
The evolution with multiplicity of the BF integral, presented in the right panel of Fig.~\ref{fig:widhtsandintegrals}, shows that with the redistribution of the charge balancing, when going to lower multiplicities, larger portions of that balancing get out of the acceptance.
With a perfect detector, $4\pi$ coverage, it should stay constant, and at one if the full momentum range were considered~\cite{Pruneau:2022mui}.

To complement the study, simulations were performed with the Pythia 8 Monte Carlo (MC) event generator at two different energies, $\sqrt{s}$ = 7 and 13.6 TeV, with the Monash tune and the ropes mode. 
\begin{figure}[hb]
    \includegraphics
    [width=0.32\textwidth,keepaspectratio=true,clip=true,trim=00pt 0pt 50pt 30pt]
    {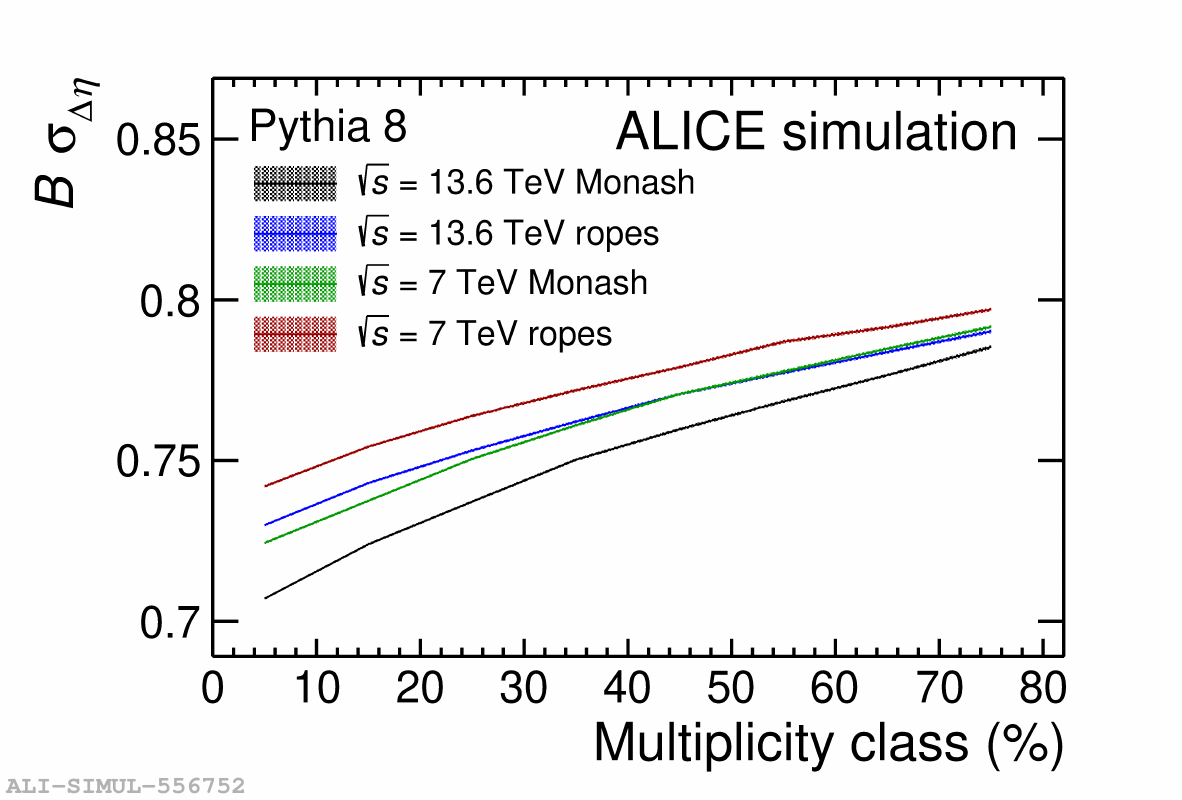}
    \includegraphics
    [width=0.32\textwidth,keepaspectratio=true,clip=true,trim=00pt 0pt 50pt 30pt]
    {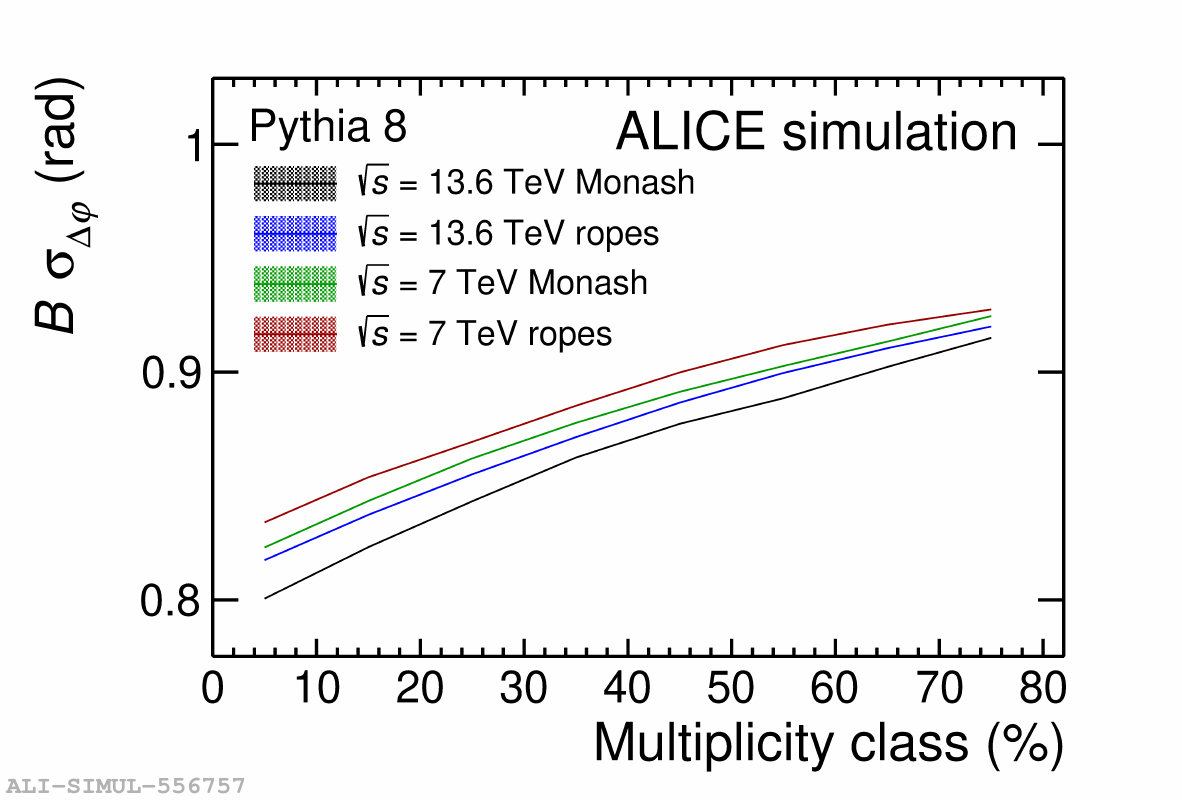}
    \includegraphics
    [width=0.32\textwidth,keepaspectratio=true,clip=true,trim=00pt 0pt 50pt 30pt]
    {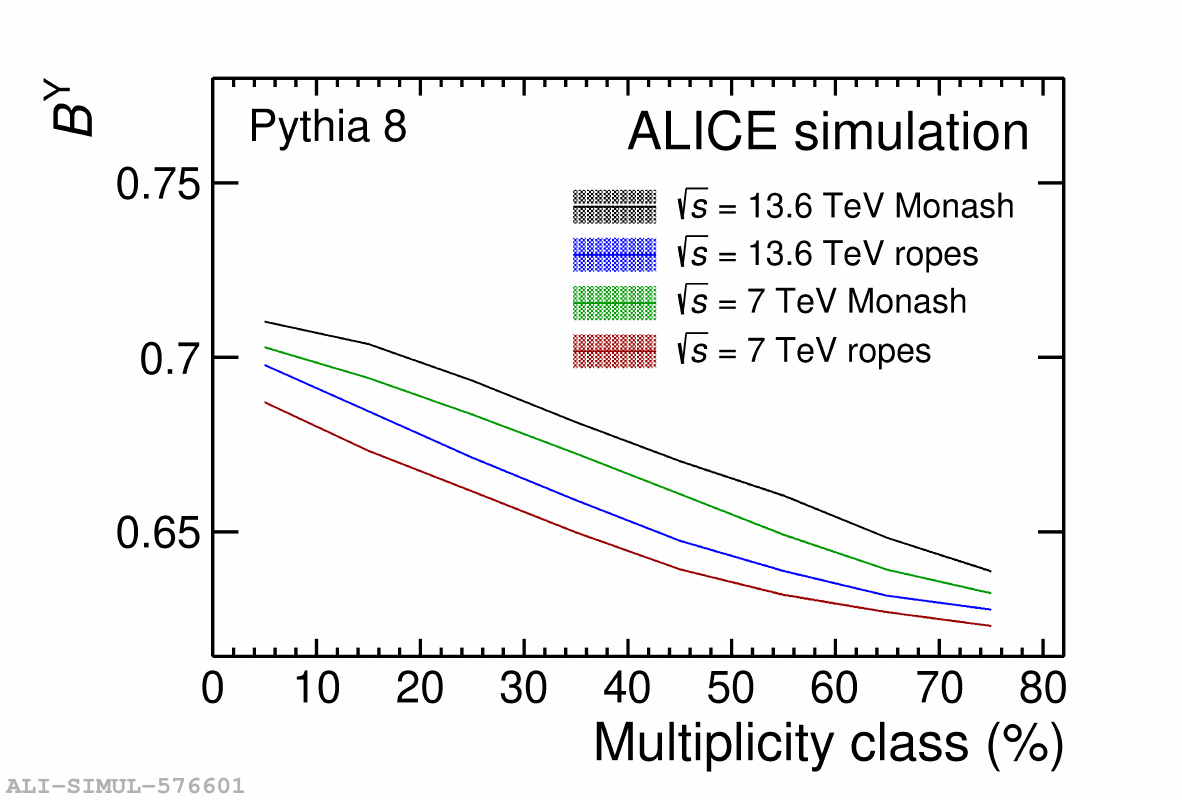}
    \caption{Evolution with multiplicity of the charged hadrons balance functions longitudinal (left) and azimuthal (center) widths, and their integrals (right), from Pythia 8 simulated pp collisions at $\sqrt{s} = 7\;\text{and}\;13.6\;\text{TeV}$ using the Monash tune and the ropes mode.}
    \label{fig:widhtsandintegralspythia}
\end{figure}
The Monash tune was developed focusing on heavy-quark fragmentation and strangeness production. Data from LHC, SPS, and Tevatron were used to constrain the initial-state-radiation, multi-parton-interaction parameters, and energy scaling. The ropes mode was developed to deal with hadronization in dense environments where overlapping strings (flux tubes) can interact to form ``ropes'' which enhance particle production, especially strange and multi-strange baryons.

The evolution with multiplicity of the charged hadrons BFs longitudinal and azimuthal widths, and their integrals, from Pythia 8 event generator are shown in Fig.~\ref{fig:widhtsandintegralspythia}. 
Qualitatively, the evolution trends shown by the data are seen as well in the simulations. 
This suggests that Pythia 8 reproduces the expansion scenario driven by increasing momentum with increased multiplicity as well as the distribution of the charge balancing when going to lowest multiplicities. Quantitatively, it seems that the ropes mode keeps the correlations slightly longer than the Monash tune, being able to better reproduce data widths values at higher multiplicities, although at low multiplicities both converge.

The drop with decreasing multiplicity, around 10\%, of the integral of the BF reproduces the behaviour of the data but the value of the integral, slightly closer to data with the ropes mode, suggests that the amount of charge balancing that Pythia 8 is able to keep within the acceptance is larger than what is observed in data. 
The fact that the amount of charge balancing within the acceptance by using the ropes mode is lower than by using the Monash tune is partially consistent with the evolution of the widths with multiplicity. This observation reinforces the fact that the correlation length observed in data is better reproduced by the ropes mode. But the fact that the 
widths are extracted only from the near side peak probably assigns to the away side part of the exceeding balancing reflected by the simulations BFs integrals.

\section{Outlook}
The evolution with multiplicity of charged hadrons BFs from pp collisions at $\sqrt{s} = 13.6\;\text{TeV}$ were presented and their main characteristics were analyzed by observing their near side longitudinal and azimuthal widths and integrals. 
They reflected a narrowing with increasing multiplicity scenario driven by the focusing effect of increasing momentum. Comparisons with Pythia 8 MC event generator suggest that the ropes mode introduces an enlarged correlation reach which seems necessary to better reproduce the evolution of data widths which is as well partially consistent with the behaviour shown by data integrals.

\begin{figure}[ht]
    \includegraphics
    [scale=0.3,keepaspectratio=true,clip=true,trim=10pt 0pt 110pt 30pt]
    {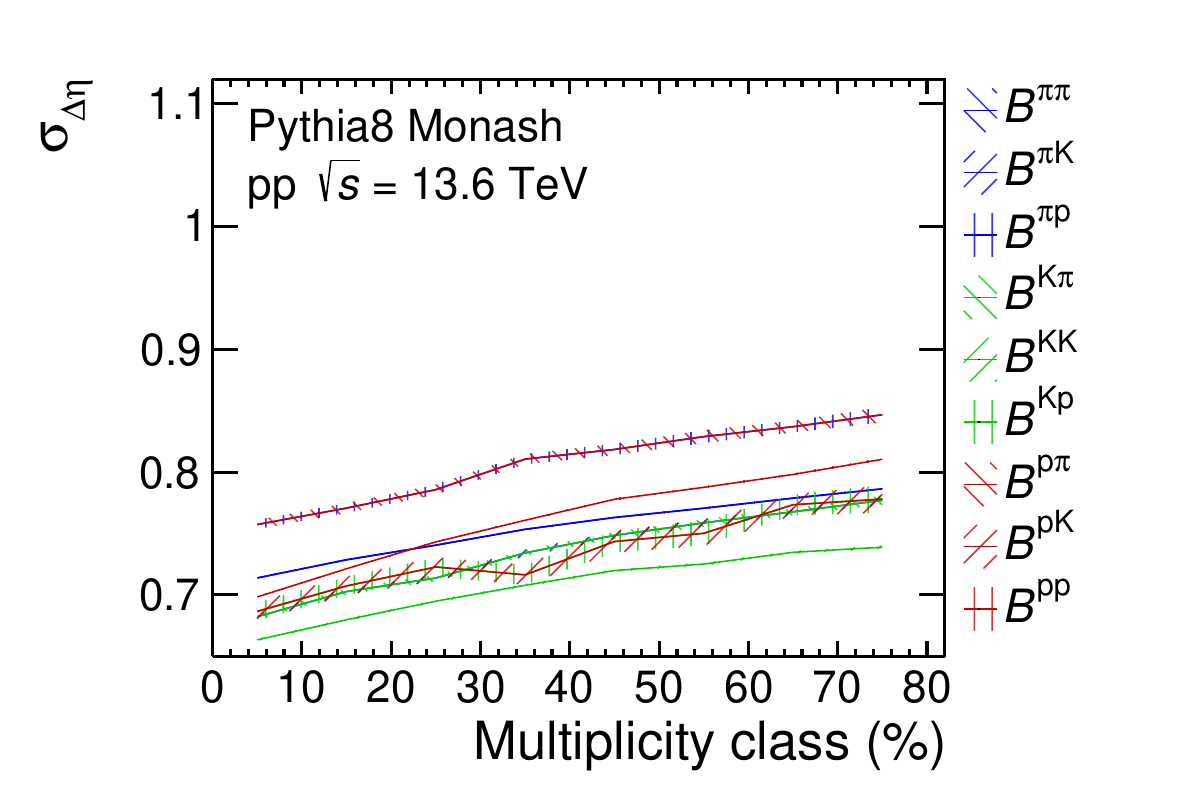}
    \includegraphics
    [scale=0.3,keepaspectratio=true,clip=true,trim=10pt 0pt 110pt 30pt]
    {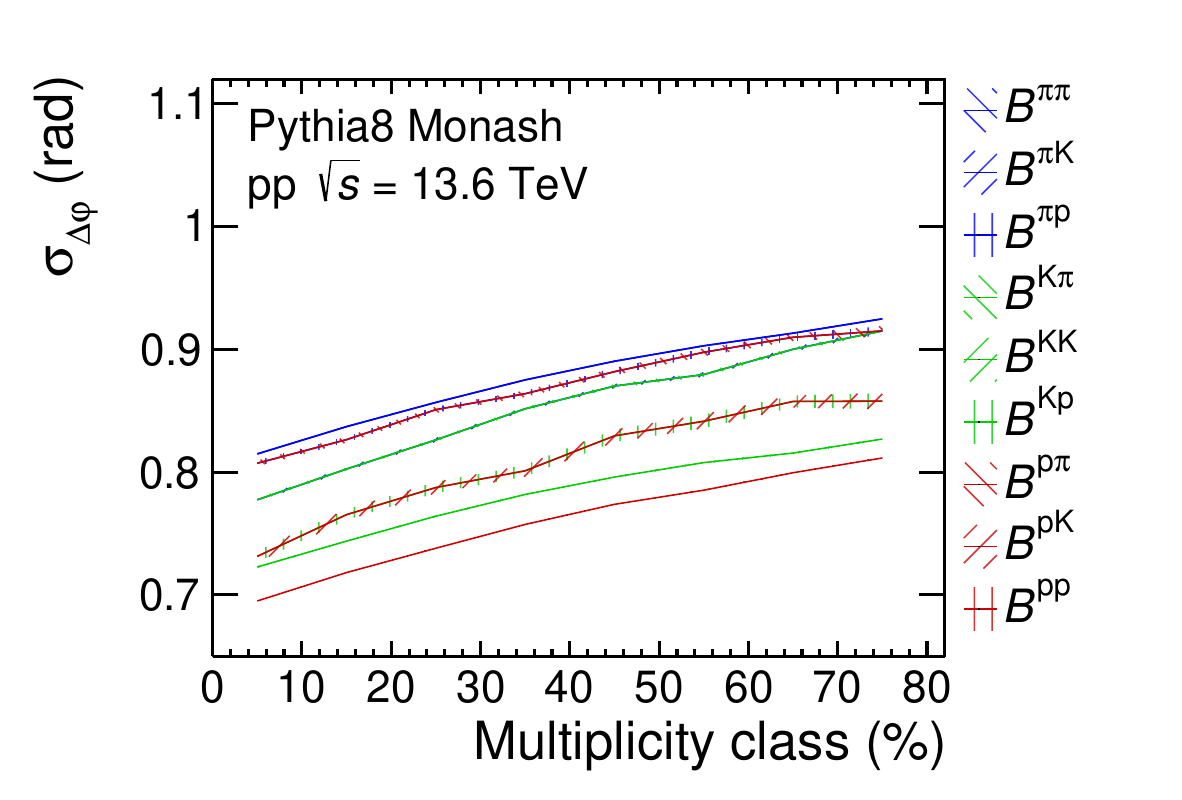}
    \includegraphics
    [scale=0.3,keepaspectratio=true,clip=true,trim=10pt 0pt 50pt 30pt]
    {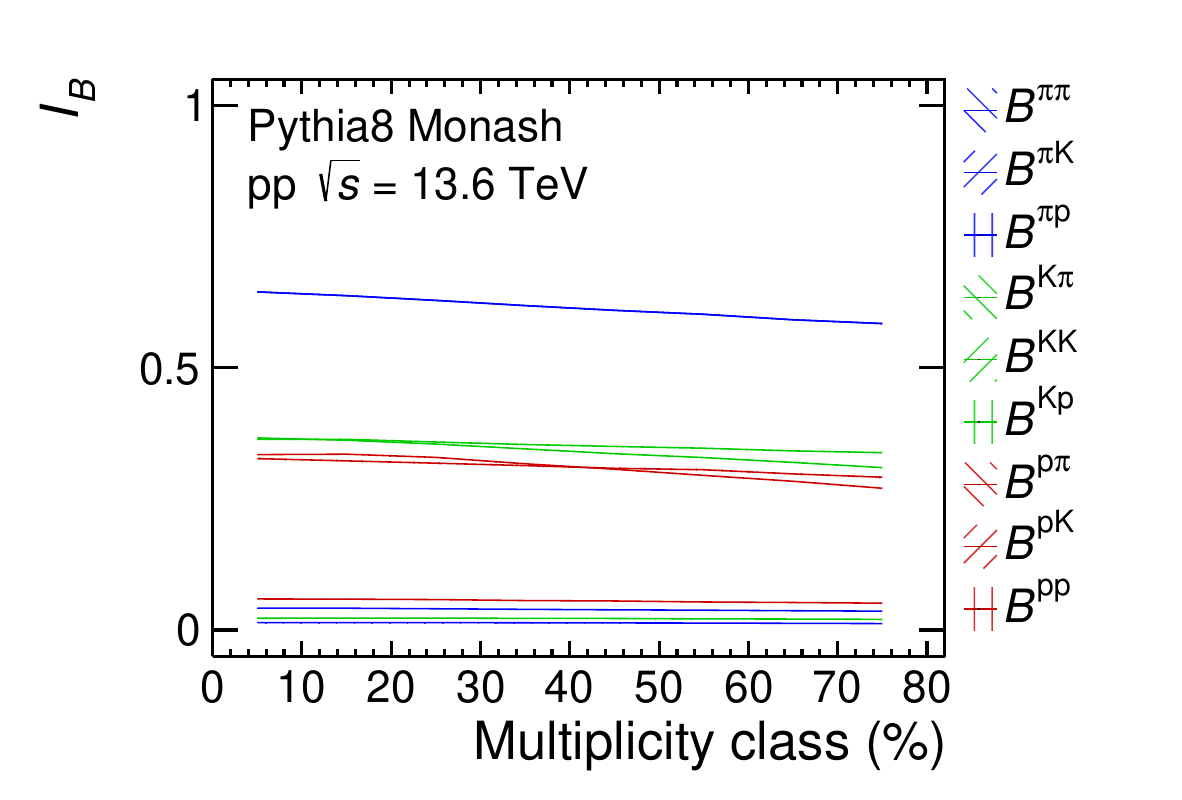}
    \caption{Evolution with multiplicity of the longitudinal (left) and azimuthal (center) widths, and integrals (right), of identified particles balance functions from Pythia 8 simulated pp collisions at $\sqrt{s} = 13.6\;\text{TeV}$ using the Monash tune.}
    \label{fig:idwidhtsandintegralspythia}
\end{figure}

The scenario is now open to address general BFs which will allow to start analyzing the balancing of, not only charge, but also baryon number and strangeness in pp collisions at LHC Run 3 energies. 
As a first step, longitudinal and azimuthal widths and integrals of BFs of identified particles from pp collisions at $\sqrt{s} = 13.6\;\text{TeV}$ produced with Pythia 8 MC event generator are shown in Fig.~\ref{fig:idwidhtsandintegralspythia}.
In principle, a narrowing with increasing multiplicity scenario is shown, with no signs of two stages of particle production and with the balancing share practically independent of multiplicity. It is worth noting that the amount of balancing of a proton by another proton, the $B^{\rm pp}$ integral, is the same if the charge is considered or if the baryon number is considered.
The same applies to strangeness. The amount of balancing of a kaon by another kaon, $B^{\rm KK}$ integral, is the same if the charge is considered or if the strangeness is considered. This depicts an scenario of interplay between the balancing of the different quantum numbers.

\bibliographystyle{JHEP}
\bibliography{bibliography}

\end{document}